# Parallel in-memory wireless computing

Cong Wang ®[1], Gong-Jie Ruan[1], Zai-Zheng Yang[1], Xing-Jian Yangdong[1], Yixiang Li ®[1] ✉, Liang Wu ®[2], Yingmeng Ge[2], Yichen Zhao ®[1], Chen Pan[3], Wei Wei[1], Li-Bo Wang ®[1], Bin Cheng ®[3], Zaichen Zhang[2], Chuan Zhang ®[2], Shi-Jun Liang ®[1] ✉ & Feng Miao ®[1,4] ✉

Parallel wireless digital communication with ultralow power consumption is critical for emerging edge technologies such as 5G and Internet of Things. However, the physical separation between digital computing units and analogue transmission units in traditional wireless technology leads to high power consumption. Here we report a parallel in-memory wireless computing scheme. The approach combines in-memory computing with wireless communication using memristive crossbar arrays. We show that the system can be used for the radio transmission of a binary stream of 480 bits with a bit error rate of 0/480. The in-memory wireless computing uses two orders of magnitude less power than conventional technology (based on digital-to-analogue and analogue-to-digital converters). We also show that the approach can be applied to acoustic and optical wireless communications.

In traditional wireless digital communication technology, the digital signal processor implements signal processing algorithms and the analogue unit transmits or receives time-varying waves carrying the digital data (such as radio waves, light waves and acoustic waves)[1,2]. These two units are physically separated[3], which gives rise to sharp interconversion interfaces between the digital and analogue domains and leads to a trade-off between power, speed and precision[4–6]. As a result, wireless digital technology faces challenges in edge device applications where ultrafast wireless communication and ultralow energy dissipation are required[7,8].

Edge device applications, thus, require alternative wireless communication technologies[9–13]. Fusing the signal processing and transmitting/receiving process units could smoothen the interconversion interfaces between the digital and analogue domains[14]. To achieve this, a hardware architecture is required that can simultaneously process wireless signals during transmission[8] such that digital data can be carried by the analogue signals and then can be extracted in an energy-efficient manner[15]. A memristive crossbar array provides a hardware platform that could implement this approach using physical laws[9,16–32].

In this article, we report a parallel in-memory wireless computing scheme. The approach is based on one-transistor–one-memristor (1T1R) crossbar arrays, which can store, transfer and modulate time-varying analogue signals and parallelly process data without digital-to-analogue converters (DACs) and analogue-to-digital converters (ADCs). We transmit 480-bit ASCII code using a transmitter and a receiver based on memristive crossbar arrays, customized peripheral circuits and radio-frequency (RF) modules. We also show that the parallel in-memory wireless computing scheme can be used for multiple-input multiple-output (MIMO) wireless communication. The approach uses two orders of magnitude less energy than wireless digital communication based on traditional technology and can be adapted to ultrasonic and optical wireless communications.

## In-memory wireless computing based on memristive crossbar arrays

Figure 1a shows the traditional digital technology and our parallel in-memory wireless computing scheme for parallel wireless digital communication. In-memory wireless computing fuses the signal processing and wireless transmission, and leverages the physical attributes of the memristive crossbar array to process the digital data carried by the time-varying analogue signals in a parallel manner[15]. Digital binary information, namely, '1' and '0', has no equivalent direct physical quantity described by the wave propagation equation and cannot be transmitted through wireless media. Therefore, binary data streams are first stored in the memristive crossbar array in the form of high- and low-resistive states. The digital data are carried by feeding carrier signals into the array and the resulting time-varying current output represents the modulated signals. The memristive crossbar architecture has parallel processing capability with multiple inputs and multiple outputs, and the modulated signals can then be transmitted through an array of antennas, piezoelectric transducers or light-emitting diodes without using DACs (Fig. 1a). The transmitted analogue signals encoding the digital data are detected by using an array of antennas, piezoelectric transducers or photodetectors.

---

[1]Institute of Brain-inspired Intelligence, National Laboratory of Solid State Microstructures, School of Physics, Collaborative Innovation Center of Advanced Microstructures, Nanjing University, Nanjing, China. [2]National Mobile Communications Research Laboratory, Southeast University and Frontiers Science Center for Mobile Information Communication and Security, Southeast University and Purple Mountain Laboratories, Nanjing, China. [3]Institute of Interdisciplinary Physical Sciences, School of Science, Nanjing University of Science and Technology, Nanjing, China. [4]Nanjing Drum Tower Hospital, The Affiliated Hospital of Nanjing University Medical School, Nanjing, China. ✉e-mail: liyx@nju.edu.cn; sjliang@nju.edu.cn; miao@nju.edu.cn

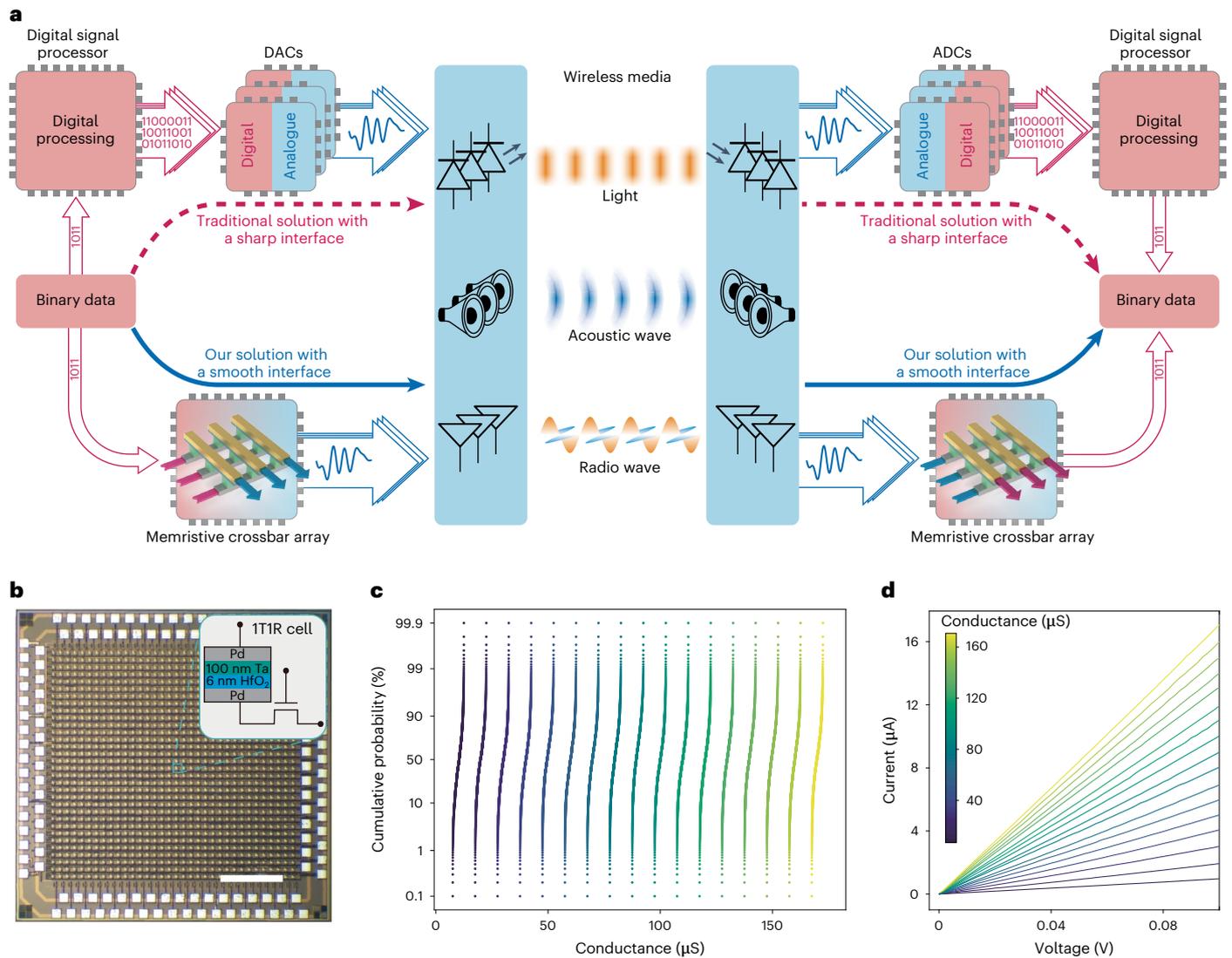

**Fig. 1 | Parallel in-memory wireless computing based on the memristive crossbar array. a**, In traditional wireless digital technology, the digital unit that implements signal processing algorithms is physically separated from the analogue transmission unit responsible for transmitting or receiving signals through wireless media such as light waves, acoustic waves and radio waves. Such a physical separation between signal processing and transmission leads to high latency and power consumption. In-memory wireless computing fuses signal processing and wireless transmission, which can eliminate the need to use an ADC or DAC. **b**, Optical photograph of a 34 × 32 1T1R memristive crossbar array. Scale bar, 500 μm. **c**, Cumulative probability distribution of programmed conductance values in 1T1R cells with respect to 17 differentiated conductance levels, each of which contains 1,000 programmed conductance values by using the closed-loop writing method. **d**, Linear current–voltage characteristics of memristive devices at different conductance states.

The programmable states of the memristive crossbar array can be used to implement a variety of complex signal processing algorithms[33–38] and thus enable the direct processing of analogue signals to extract binary information without utilizing ADCs. Fusing in-memory computing based on the memristive crossbar with wireless digital communication in this way can realize parallel in-memory wireless computing and smoothen the sharp interconversion interfaces between the digital and analogue domains; it also has the potential to eventually eliminate the trade-off issues associated with DACs and ADCs.

Figure 1b demonstrates a 34 × 32 1T1R crossbar array in which the transistor is used to switch the current on and off through the memristive device. We use a closed-loop writing method (Supplementary Fig. 1) to program the memristive devices 17,000 times. The resulting 17,000 conductance values are distributed across 17 different conductance ranges and correspond to 17 levels (Fig. 1c). These different conductance levels are well separated and have no overlap, which is required for reliable wireless communication using our 1T1R cells (Methods and Supplementary Figs. 1–3 provide a more detailed discussion on the effect of state variability). Our 1T1R cell also shows linear current–voltage characteristics (Fig. 1d).

Figure 2a shows the schematic of the circuit of a transmitter that uses a 1T1R memristor crossbar array and applies in-memory wireless computing. The crossbar array not only stores the transferred digital data but can also modulate the signal. We employ orthogonal frequency-division multiplexing (OFDM) and quadrature amplitude modulation (QAM), widely used in 5G mobile communication, as a proof-of-concept demonstration. Orthogonal signals (that is, $\cos(\omega t)$, $-\sin(\omega t)$, $\cos(2\omega t)$, $-\sin(2\omega t)$…) are fed into the memristor crossbar array in parallel, and the current outputs from the array represent the modulated signals. Note that the orthogonal subcarrier signals can also be generated by using the memristive crossbar array (Supplementary Fig. 4). Since the adjacent columns in the memristive crossbar array are used to map the positive and negative conductance values, the modulated signal is the difference between the current output from the even-numbered column and that of the odd-numbered column. To transmit more digital data, the symbols are sequentially changed by switching the transistors on the columns of the array on and off, and the transmitted data in each symbol period can be identified by selected memristive devices. The output from the memristor crossbar array is upconverted to the high-frequency passband, amplified and eventually transmitted into free space by an antenna.

We used QAM and experimentally mapped 480-bit ASCII codes corresponding to the character string 'Here is a demo of memristor-based communication system @ NJU' into the conductance states of a 30 × 32 memristive crossbar array (Fig. 2b). Using 30 orthogonal subcarrier sources supporting 15-OFDM; customized peripheral circuits; and RF modules including amplifier, mixer and antenna (Methods provides more details about the RF module), we show the wireless digital transmission of 480-bit data. Figure 2c shows the baseband signals and passband signals within two symbol periods. The signals were measured in each symbol period using an oscilloscope as the transistors connected in series with the memristors of two adjacent columns were switched on. The constellation diagram of the transmitted digital data (Fig. 2d) shows that the DAC can be removed when using in-memory wireless computing for wireless digital transmission.

In-memory wireless computing can also be used to extract digital data from transmitted wireless analogue signals with a bit error rate of 0. Figure 2e shows the schematic of the circuit of a receiver based on the memristive crossbar array. The transmitted signals are received by an antenna and undergo amplification, downconversion and sampling (Methods). The resulting signals from the RF module are then sequentially fed into each row of the memristive crossbar array using the multiplexer. The negative weight is realized through one-resistor-one-memristor differential scheme. The current output from each column is integrated within a symbol period (from $t_1$ to $t_{32}$). The signals obtained within each symbol period, representing the in-phase components and quadrature components at 15 frequencies of the received signal, can be binarized by comparators to rebuild the transmitted digital data. Figure 2f,g shows the wireless signals received by the RF module in the receiver and the conductance matrix used to implement the discrete Fourier transform (DFT) algorithm with the memristive crossbar array, respectively. The constellation diagram (Fig. 2h) shows that the transmitted digital data were successfully extracted from the received wireless signals without using the ADC.

Other signal processing algorithms in conjunction with the demodulation algorithm, such as windowing and channel equalization, can also be implemented using the same memristive crossbar array, without consuming more energy and hardware resources (Methods). Our proposed in-memory wireless computing can be readily extended to high-order modulation (Supplementary Fig. 5) and can be combined with other mature channel coding technologies, such as Hamming codes, low-density parity-check codes, turbo codes and polar codes, to further improve the system performance in practical applications.

## Parallel in-memory wireless computing for MIMO systems

Combining in-memory wireless computing with MIMO technology has the potential for the parallel processing of wireless signals through multipath propagation, which can enhance the capacity of simultaneous data streams[39]. Figure 3a shows a schematic of MIMO wireless communication based on parallel in-memory wireless computing. Multiple OFDM signals ($\mathbf{Tx}_{1,2…N}$) carrying different binary data can be simultaneously generated by the memristive crossbar array and transmitted through an array of antennas. After multipath propagation, the resulting signals ($\mathbf{Rx}_{1,2…M}$) are received by an array of antennas in the receiver, according to the MIMO system model at the $k$th subcarrier (given as $\mathbf{Rx}(k) = H(k)\mathbf{Tx}(k) + \mathbf{n}(k)$ (ref. [40]), where $H(k)$ is the channel matrix of the $k$th subfrequency channel and $\mathbf{n}(k)$ is the noise in the $k$th subfrequency channel). To extract the binary data in parallel, the received signals undergoing multipath propagation are simultaneously input into a memristive crossbar array and processed using time-to-frequency transformation and MIMO decoupling algorithms. By taking advantage of the parallel architecture of the memristive crossbar array, we can combine the weight matrix $T_f$ of the time-to-frequency transformation algorithm with matrix $W$ of the MIMO decoupling algorithm into a weight matrix $G$ in the memristive crossbar array, which is given as

$$G = \begin{bmatrix} f(T_f, W_{11}) & \cdots & f(T_f, W_{1N}) \\ \vdots & \ddots & \vdots \\ f(T_f, W_{M1}) & \cdots & f(T_f, W_{MN}) \end{bmatrix}, \quad (1)$$

where $f(T_f, W_{MN}) = [T_f(1) \times W_{MN}(1) … T_f(k) \times W_{MN}(k)]$, $T_f(k) = \begin{bmatrix} e^{-i2\pi k(0/L)} \\ e^{-i2\pi k(1/L)} \\ \vdots \\ e^{-i2\pi k(L-1)/L} \end{bmatrix}$

$k$ represents the number of subfrequency channels and $L$ is the length of DFT. By adopting the zero-forcing precoding approach, the weight matrix $W(k)$ used for implementing the MIMO decoupling algorithm is determined by either $H(k)^{-1}$ or $(H(k)^H H(k))^{-1} H(k)^H$, for $N = M$ or $N < M$. We verify the parallel in-memory wireless computing scheme with a 2 × 2 MIMO system using 7-OFDM and 4-QAM (Fig. 3b–e). The obtained results indicate that the memristive crossbar arrays can be used to build parallel wireless digital communication systems without employing a DAC and ADC. Furthermore, parallel complex-valued matrix–vector multiplication can be realized on the crossbar array (Supplementary Fig. 6).

**Fig. 2 | In-memory wireless computing by using a memristive crossbar array. a**, Schematic of the circuit of a DAC-free wireless transmitter based on a memristive crossbar array. Orthogonal signals are fed into a memristive crossbar array in parallel, and an OFDM baseband signal is generated in the form of an output current at selected columns. The output signal is upconverted to a high frequency and transmitted by an antenna. LO, local oscillator; AMP, amplifier. **b**, The 480-bit data are encoded in the conductance matrix, which represent the character string 'Here is a demo of memristor-based communication system @ NJU' in ASCII code. **c**, The generated baseband signals and passband signals of the transmitted symbols at different symbol periods ($T_1$ and $T_2$). **d**, Constellation diagram of all the transmitted signals. The abscissas and ordinates of the points represent the in-phase (I) and quadrature (Q) components at all the subcarrier frequencies, respectively. **e**, Schematic of the circuit of the ADC-free wireless receiver based on a memristive crossbar array. **f**, A downconverted signal is sequentially applied on the 32 different electrodes of the memristive crossbar array from $t_1$ to $t_{32}$. **g**, A conductance matrix is mapped into a memristive crossbar array for implementing the DFT algorithm and other algorithms including windowing and channel equalization. **h**, Measured constellation diagram of all the received signals.

## Ultrasonic and optical wireless communications

Our parallel in-memory wireless computing technology is not limited to radio waves and can be applied to other transmission media. Figure 4a–f demonstrates wireless digital transmission by ultrasonic waves and light waves in which piezoelectric transducers are used for transmitting and detecting ultrasonic waves, whereas light-emitting diodes and photodiodes are used to transmit and detect the light intensity. We utilize a memristor crossbar array to transform the digital data into baseband signals and the generated modulation signals from the array are upconverted and transmitted through piezoelectric transducers or light-emitting diodes, and the signals received by piezoelectric transducers or photodiodes (Fig. 4b,e, middle) are downconverted and demodulated using the memristive crossbar array to extract the binary data. Figure 4c,f demonstrates the constellation diagram of the received digital data using ultrasonic and optical wireless communications, respectively. The received data are the same as those stored in the memristive crossbar array of the transmitter.

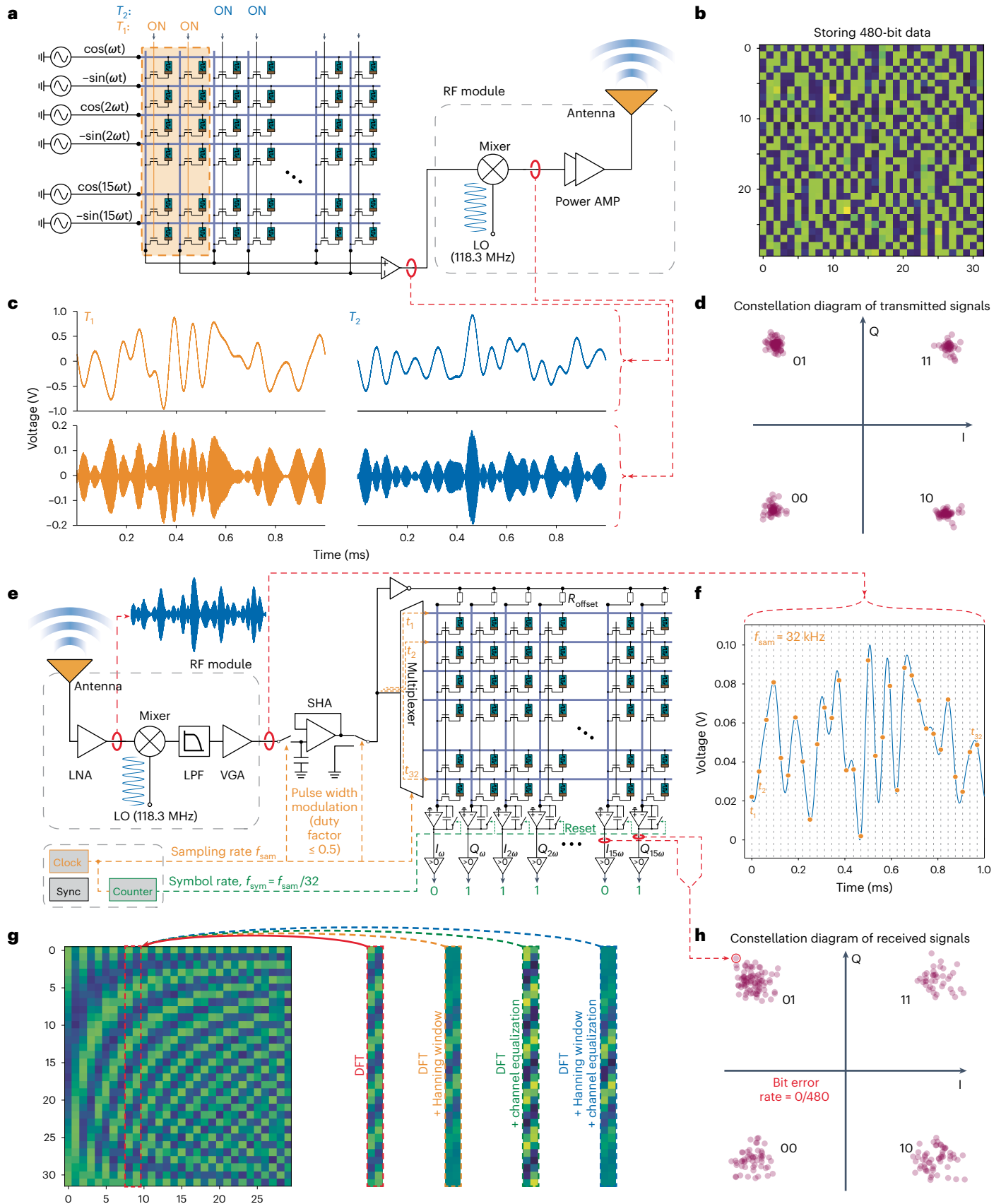

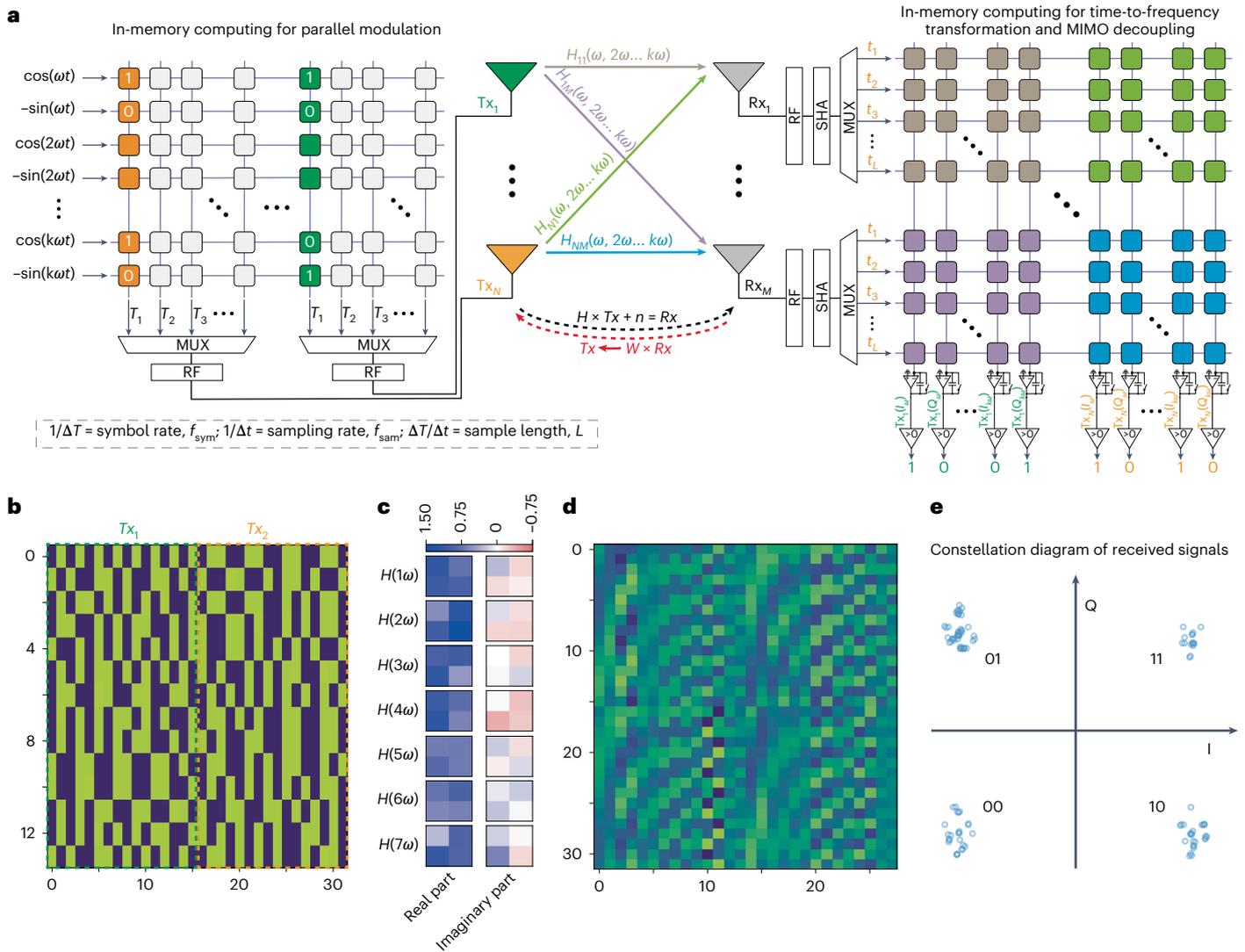

**Fig. 3 | Parallel in-memory wireless computing. a**, MIMO wireless communication based on parallel in-memory wireless computing. MUX, multiplexer. **b**, Weight matrix of the left crossbar array. **c**, The 2 × 2 complex-valued channel matrices at different frequencies. **d**, Weight matrix of the right crossbar array. The DFT and MIMO decoupling operations were fused and mapped onto a single crossbar array. **e**, Constellation diagram of the received signals, showing that all the 224-bit data can be successfully extracted from the received signals with a bit error rate of 0.

## Bit error rate and energy efficiency

We evaluate the state variation dependence of the bit error rate by simulating the data transmission process (Fig. 2a,e) (Methods provides more simulation details) and plot the results (Fig. 4g). The simulation result shows that the bit error rate is fairly small even if the programming error of the memristive devices is 25%. Compared with the programming error used in the simulation, the programming error of our fabricated 1T1R cells is one order of magnitude lower than 25%. Supplementary Fig. 1 shows a programming error of 1.18%, indicating that the wireless transmission system based on the in-memory wireless computing can be reliably operated.

We compared the energy consumption for both digitalization and signal processing of our scheme with the traditional digital solution using ADCs. We consider scenarios with the same signal-to-noise ratio (Methods and Supplementary Fig. 7). In traditional wireless digital transmission, the digitalization process takes place before signal processing, whereas the equivalent digitalization process in our scheme is shifted backwards. Therefore, the energy consumption associated with digitalization can be reduced by 1/140 in our scheme compared with the traditional wireless digital transmission based on the ADC (Fig. 4h). Methods and Supplementary Figs. 7 and 8 provide more details for these calculations. The energy efficiency for in-memory wireless computing is two orders of magnitude higher than digital signal processing (Fig. 4i and Methods). Note that prior work has demonstrated that the memristive devices can be used as non-volatile RF switches for 6G communication systems[12]. In conjunction with the frequency characteristic and efficiency performance, in-memory wireless computing based on the memristive crossbar array has the potential for wireless data transmission and networking systems that can be deployed in edge devices.

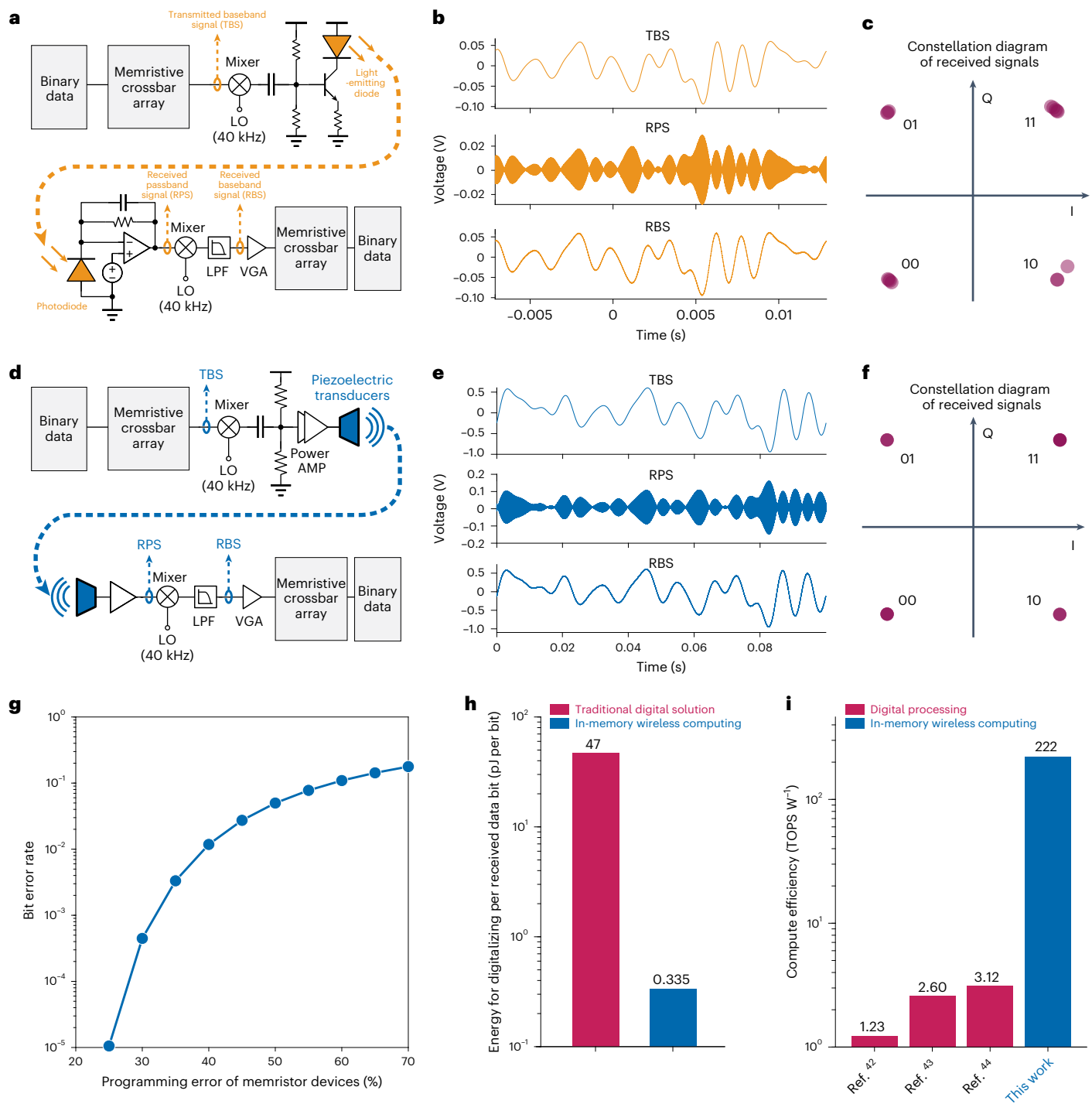

**Fig. 4 | Ultrasonic and optical wireless communications, bit error rate and energy efficiency. a**, Schematic of the experimental circuit for optical wireless communication. **b**, Measured signal waveforms at different transmission stages. **c**, Constellation diagram of the received signals. **d**, Schematic of the experimental circuit for ultrasonic wireless communication. **e**, Measured signal waveforms at different transmission stages. **f**, Constellation diagram of the received signals. **g**, State variation dependence of the bit error rate, which is characterized by the programming error of the memristive devices. **h**, Energy consumptions for digitalizing the analogue signals in in-memory wireless computing and the ADC-based traditional solution. **i**, Energy efficiency of signal processing based on in-memory wireless computing and traditional digital processing[42–44]. TOPS W$^{-1}$, tera-operations per second per watt.

## Conclusions

We have reported a parallel in-memory wireless computing scheme based on memristive crossbar arrays. We showed that the approach can be used for the wireless radio transmission of 480-bit binary data with a bit error rate of 0 and can be applied to other transmission media (ultrasonic waves and light waves). We also showed that our parallel wireless digital communication technology uses around two orders of magnitude less energy than conventional technology.

## Methods

### 1T1R device fabrication and integration

We fabricated the 1T1R crossbar array chip by integrating the Pd/Ta/HfO$_2$/Pd memristive devices onto a custom complementary metal–oxide–semiconductor chip. On-chip multiplexers, decoders, transistor arrays and metal interconnections were fabricated in a standard complementary metal–oxide–semiconductor foundry based on the 180 nm technology node. The memristive devices were fabricated in the laboratory. The memristive devices were integrated onto the chip by using standard photolithography, etching and deposition processes. The bottom electrode layer of Pd (50 nm thick) was deposited by using electron-beam deposition; the switching layer of HfO$_2$ (6 nm thick), by atomic layer deposition; the capping layer of Ta (100 nm thick), through a standard RF sputtering process in Ar; and the top electrode layer of Pd (50 nm thick), using electron-beam deposition. All the above layers were patterned by standard photolithography and were lifted off in acetone. The architecture of the 1T1R chip is shown in Supplementary Fig. 9.

### OFDM–QAM modulation

To implement OFDM–QAM modulation, orthogonal signals (cos(1,000 × 2π × $t$), −sin(1,000 × 2π × $t$), cos(2,000 × 2π × $t$), −sin(2,000 × 2π × $t$)…cos(15,000 × 2π × $t$), −sin(15,000 × 2π × $t$)) were fed into the memristive crossbar array, weighted by the conductance of the memristive devices and then accumulated. Since the positive and negative conductance values are mapped into adjacent columns of the memristive crossbar array, the current output from the even-numbered columns was subtracted from that of the odd-numbered columns by a differential current-to-voltage amplifier. The transmitted symbol can be sequentially changed by switching the transistors on two adjacent columns on and off, and the transmitted data in each symbol period are identified by the selected memristive devices. As a result, current outputs generated from the array represent the baseband signal of an OFDM symbol. The output signal from the memristive crossbar array was upconverted to a high frequency (118.3 MHz) and then transmitted by an antenna. The waveforms at different stages were measured by an oscilloscope. The local oscillating signal was generated by a Keysight 81150A generator.

### OFDM–QAM demodulation

The wireless signal was received by an antenna and amplified by a low-noise amplifier, which was eventually downconverted to a low frequency by a mixer and low-pass filter (LPF). Considering that the magnitude of the wireless signal is reduced by various environmental factors, a variable-gain amplifier (VGA) has been used to adjust the signal amplitude to an appropriate range, which was sequentially sampled by a sample-and-hold amplifier (SHA) with a sampling rate $f_{sam}$ (32 kHz) determined by the switching frequency of the multiplexer. In a symbol period (1 ms), the 32 sampled signals were sequentially applied on 32 rows of the memristive crossbar array with a multiplexer. Experimentally, we realized the negative weight using a resistor that generates an offset current through $R_{offset}$ at each column. For a memristive crossbar array to implement DFT, the conductance value of the memristive device at the $i$th row and $j$th column is determined by $A \times \cos(2\pi \times j \times (i-1)/32) + G_{offset}$ or $A \times \sin(-2\pi \times j \times (i-1)/32) + G_{offset}$ according to $j = 2k-1$ or $2k$, where $k = 1, 2…15$. The output current was accumulated by an integrator in a symbol period, and the resulting voltages from the integrators represent the in-phase components and quadrature components of the received symbol, which are finally binarized by using comparators. Note that the proposed scheme is fully compatible with a higher-dimensional OFDM system due to the excellent scalability of the memristive crossbar array.

### Radio, optical and acoustic wireless communications

The radio, optical and acoustic transmitting and receiving modules in the experiments were designed and assembled by using customized printed circuit boards. The schematic of the transmitting and receiving modules in radio wireless communication is shown in Fig. 2a,e, and the corresponding ones for optical and acoustic wireless communications are presented in Fig. 4a,d, respectively. In the experimental demonstration, local oscillating signals were generated by a Keysight 81150A generator, signals were measured by a Tektronix MSO54 oscilloscope and other parts were realized by commercial devices (for example, AD835 mixer, OPT101 photodetector and OPA192 amplifier).

### Conductance programming of memristive devices and variability analysis

We experimentally demonstrated 17 different conductance levels in our fabricated 1T1R cells and used them for the proof-of-concept demonstration of the in-memory wireless computing (Fig. 1c). To reliably differentiate these conductance states, we gradually programmed the conductance values of the Ta/HfO$_2$ memristive devices by applying voltage pulses through an Agilent B1530 pulse generator. Accordingly, we programmed the memristor conductance into a defined range from $G_{target} - \Delta G$ to $G_{target} + \Delta G$ using a closed-loop writing method implemented with a script in LabVIEW 2010: if $G_{read} < (G_{target} - \Delta G)$, a positive pulse would be applied on the memristive device; else if $G_{read} > (G_{target} + \Delta G)$, a negative pulse would be applied. The experimental results demonstrate that all the programmed memristive devices are located within the defined conductance range (Supplementary Fig. 1). The deviations of the programmed conductance values are uniformly distributed from −2.5 to 2.5 μS (Supplementary Fig. 1d), corresponding to a variation of about 5.0 μS for each conductance level. Note that the conductance variation can be further reduced below 2 μS (Supplementary Fig. 1f,g) by choosing appropriate initial parameters for $\Delta G$ used in the closed-loop writing method (Supplementary Fig. 1a). Such a small variation in the conductance level corresponds to a programming error ranging from 1.18% to 2.94%, which is defined by $2\Delta G/(G_{max} - G_{min})$ (Supplementary Fig. 1b), where $G_{max}$ and $G_{min}$ represent the maximum and minimum values within the continuously tuned conductance range of the memristive devices, respectively.

### Energy efficiency analysis

Using the memristive crossbar array to process wireless signals can significantly reduce the energy consumption associated with the ADCs in a traditional receiver, because simple comparators can be used for replacing ADCs to digitalize the analogue signals in our work. To illustrate the advantage of in-memory wireless computing in energy efficiency, the power consumption of the ADCs and comparators was calculated for an OFDM–4-QAM wireless communication system. For the ADC design, there is a general trade-off between the resolution and power efficiency. To identify the trends and limits, we collected the data of the ADC designs published in ISSCC and VLSI Circuit Symposium from 1999 to 2021 (ref. 41) (Supplementary Fig. 8). The result shows that the energy consumption per sample per bit of the ADC exponentially increases with the corresponding effective number of bits (ENOB). The relation between power consumption and ENOB can be well described by $y = 10^{0.2315x - 0.7068}$, where $y$ is the power consumption and $x$ is the ENOB. Such an equation reveals a fact that replacing high-precision ADCs with low-precision ones (even comparators) can significantly reduce the energy consumption per sample per bit.

As shown in Supplementary Fig. 7a, for an OFDM–4-QAM wireless communication system, the ADC is employed in a traditional digital receiver for digitalizing the received analogue signals before processing the signal in the digital domain, with high resolution and high sampling rate $f_{sam}$ (equal to $\alpha$ × number of subcarriers × symbol rate, where factor $\alpha$ has to be greater than 2 owing to Nyquist's theorem). In contrast, the comparator with a 1-bit ENOB was used in our work for digitalizing the processed analogue signals, with the sampling rate equal to symbol rate $f_{sym}$. Considering that our scheme that uses a memristive crossbar array with 1.18% programming error should have an equivalent signal-to-noise ratio as the traditional solution using ADCs with 6.7-bit ENOB (Supplementary Fig. 7b), the reduction in power consumption by using in-memory wireless computing can be evaluated by the following equation:

$$\frac{6.7 \times \text{ADC\_number} \times f_{\text{sam}} \times \text{ADC\_Energy\_per\_sample\_per\_bit}}{1 \times \text{Comparator\_number} \times f_{\text{sym}} \times \text{Comparator\_Energy\_per\_sample\_per\_bit}}$$

where $\frac{\text{ADC\_Energy\_per\_sample\_per\_bit}}{\text{Comparator\_Energy\_per\_sample\_per\_bit}}$ can be calculated to be 20.9 by using the fitted equation $y = 10^{0.2315x - 0.7068}$; the minimum value of $\frac{\text{ADC\_number} \times f_{\text{sam}}}{\text{Comparator\_number} \times f_{\text{sym}}}$ can be calculated to be 1 according to Nyquist's theorem. Based on these parameters, 140 is the value obtained for the reduction in energy consumption for digitalizing analogue signals in wireless communication.

**Computing efficiency based on the memristive crossbar array**

In our proposed architecture, the signal processing was realized by an end-to-end analogue in-memory wireless computing scheme. For the architecture presented in Fig. 2e, the energy consumption per second on memristive devices and offset resistors can be calculated by using the following equation: (equivalent voltage)² × (averaged memristor conductance + conductance of $R_{\text{offset}}$) × duty factor × 2 × number of subcarriers. The computation per second is equal to 2 × number of subcarriers × sampling rate. Hence, the computing efficiency can be calculated by using the following equation: computation per second/energy consumption per second = sampling rate/(equivalent voltage)²/(averaged memristor conductance + conductance of $R_{\text{offset}}$)/duty factor = 100 MHz × 2/(0.1 V)²/(90 + 90 µS)/0.5 = 222 tera-operations per second per watt.

## Data availability

Source data are provided with this paper. All other data that support the plots within this paper and other findings of this study are available from the corresponding authors upon reasonable request.

## Acknowledgements
This work was supported in part by the National Natural Science Foundation of China (62122036, 62034004, 61921005 and 61974176) and the Strategic Priority Research Program of the Chinese Academy of Sciences (XDB44000000). F.M. would like to acknowledge support from the AIQ Foundation. The microfabrication center of the National Laboratory of Solid State Microstructures (NLSSM) is acknowledged for their technique support.

## Author contributions
C.W. and G.-J.R. equally contributed to this work. F.M., S.-J.L. and C.W. conceived the idea and designed the experiments. F.M. and S.-J.L. supervised the whole project. G.-J.R. and C.W. performed all the experiments and analysed the experimental data. Z.-Z.Y., Y.Z. and Y.L. aided in the experiment design. Z.-Z.Y., W.W. and X.-J.Y. provided help in the device fabrication and circuit assembly. C.P., L.-B.W. and B.C. contributed to the circuit measurements. L.W., C.Z., Y.G. and Z.Z. contributed to the modelling of the wireless communication systems and the corresponding experimental design. C.W., S.-J.L. and F.M. co-wrote the paper.

## Competing interests
The authors declare no competing interests.

## Additional information
**Supplementary information** The online version contains supplementary material available at https://doi.org/10.1038/s41928-023-00965-5.

**Correspondence and requests for materials** should be addressed to Yixiang Li, Shi-Jun Liang or Feng Miao.

**Peer review information** *Nature Electronics* thanks Howard Hao Yang, Basem Shihada and the other, anonymous, reviewer(s) for their contribution to the peer review of this work.

**Reprints and permissions information** is available at www.nature.com/reprints.

**Publisher's note** Springer Nature remains neutral with regard to jurisdictional claims in published maps and institutional affiliations.